\documentstyle{article}[12pt]
\def \lket {|}
\def \rket {\rangle}
\def \lbra {\langle}
\def \rbra {|}
\def \H{{\cal H}}
\newcommand{\ket}[1]{\lket #1\rket}
\newcommand{\bra}[1]{\lbra #1\rbra}

\newtheorem{Theorem}{Theorem}
\newtheorem{Lemma}{Lemma}

\newcommand{\proof}{\noindent {\bf Proof: }}
\newcommand{\qed}{$\Box$}
\newcommand\comment[1]{}

\begin{document}
\title{Quantum lower bounds by quantum arguments}

\author{Andris Ambainis\\
       Computer Science Division\\
       University of California\\
       Berkeley, CA 94720\\
       e-mail: ambainis@cs.berkeley.edu\thanks{Supported
       by Berkeley Fellowship for Graduate Studies and, in part,
       by NSF grant CCR-9800024. This research was done while
       visiting Microsoft Research.}}
\date{}

\maketitle

\begin{abstract}
We propose a new method for proving lower bounds on quantum query algorithms.
Instead of a classical adversary that runs the algorithm with one input
and then modifies the input, we use a quantum adversary that runs
the algorithm with a superposition of inputs. 
If the algorithm works correctly, its state becomes entangled with
the superposition over inputs. 
We bound the number of queries needed to achieve a sufficient entanglement
and this implies a lower bound on the number of queries
for the computation.

Using this method,
we prove two new $\Omega(\sqrt{N})$ lower bounds on computing AND of ORs
and inverting a permutation and also provide more uniform proofs 
for several known lower bounds which have been previously proven via
variety of different techniques.
\end{abstract}

\section{Introduction}

In the query model, algorithms access the input only by querying 
input items and the complexity of the algorithm is measured by the number
of queries that it makes. Many quantum algorithms can be naturally expressed 
in this model. The most famous examples are Grover's
algorithm\cite{Grover} for searching an $N$-element list with 
$O(\sqrt{N})$ quantum queries and period-finding which is the basis
of Shor's factoring algorithm\cite{Kitaev,Shor}.

In the query setting, one can not only construct efficient quantum
algorithms but also prove lower bounds on the 
number of queries that any quantum
algorithm needs. For example, it can be shown that any algorithm 
solving the unordered search problem 
needs $\Omega(\sqrt{N})$ queries\cite{BBBV}.
(This implies that Grover's algorithm is optimal.)

The lower bounds in quantum query model provide insights into
the limitations of quantum computing.
For example, the unordered search problem provides an
abstract model for NP-complete problems and the $\Omega(\sqrt{N})$ lower
bound of \cite{BBBV} provided evidence of the difficulty of
solving these problems on a quantum computer.

For two related problems - inverting a permutation (often used
to model one-way permutation) and AND of ORs
only weaker lower bounds have been known.
Both of these problems can be solved using Grover's algorithm
with $O(\sqrt{N})$ queries for inverting a permutation and
$O(\sqrt{N} \log N)$ queries for AND of ORs\cite{BCW}.
However, the best lower bounds have been $\Omega(\sqrt[3]{N})$\cite{BBBV}
and $\Omega(\sqrt[4]{N})$, respectively.  

We present a new method for proving lower bounds on quantum query algorithms
and use it to prove $\Omega(\sqrt{N})$ lower bounds for inverting a 
permutation and AND of ORs. 
It also provides a unified proof for several other results
that have been previously proven via variety of different techniques.

In contrast to \cite{BBBV,V} that use a classical adversary argument
(an adversary runs the algorithm with one input and, after that, changes
the input slightly so that the correct answer changes but the algorithm
does not notice that), we use a quantum adversary.
In other words, instead of running the algorithm with one input,
we run it with a superposition of inputs.
This gives stronger bounds and can also simplify the proofs.

More formally, we consider a bipartite quantum system consisting of the
algorithm and an oracle answering algorithm's queries. 
At the beginning, the algorithm part is
in its starting state (normally $\ket{0}$), the oracle part
is in a uniform superposition over some set of inputs and the two parts
are not entangled.

In the query model, the algorithm can either perform a unitary transformation
that does not depend on the input or a query transformation 
that accesses the input.
The unitary transformations of the first type become unitary transformations
over the algorithm part of the superposition. 
The queries become transformations
entangling the algorithm part with the oracle part.

If the algorithm works correctly,
the algorithm part becomes entangled with the oracle part
because the algorithm part must contain different answers
for different inputs.
We obtain lower
bounds on quantum algorithms by bounding the number of
query transformations needed to achieve such entanglement.


Previously, two main lower bound methods were 
classical adversary\cite{BBBV} (called 'hybrid argument' in \cite{V}) 
and polynomials methods.
The classical adversary/hybrid method of \cite{BBBV,V} 
starts with running the algorithm on one input. 
Then the input is modified so that the behavior of algorithm
does not change much but the correct answer does change. 
That implies that the problem cannot be solved with a small number of 
queries. 
Polynomials method\cite{Beals} uses the fact that 
any function computable with a small number of queries can be
approximated by a polynomial of a small degree and then applies
results about inapproximability by polynomials.

Our ``quantum adversary'' method can be used to give more unified
proofs for many (but not all) results that were previously shown 
using different variants of hybrid and/or polynomials method.


There is also a new proof of the $\Omega(\sqrt{N})$ lower bound 
on unordered search by Grover\cite{Grover1}. 
This proof is based on considering the sum of distances between
superpositions on different inputs.
While the motivation for Grover's proof (sum of distances)
is fairly different from ours (quantum adversary), these
two methods are, in fact, closely related.
We discuss this relation in section \ref{S7}.

\section{The model}

We consider computing a Boolean function
$f(x_1, \ldots, x_N):\{0, 1\}^N\rightarrow \{0, 1\}$
in the quantum query model\cite{Beals}.
In this model, the input bits can be accessed by queries to an oracle $X$
and the complexity of $f$ is the number of queries needed to compute $f$.

A quantum computation with $T$ queries
is just a sequence of unitary transformations
\[ U_0\rightarrow O\rightarrow U_1\rightarrow O\rightarrow\ldots
\rightarrow U_{T-1}\rightarrow O\rightarrow U_T.\]

$U_j$'s can be arbitrary unitary transformations that do not depend
on the input bits $x_1, \ldots, x_N$. $O$ are query (oracle) transformations.
To define $O$, we represent basis states as $|i, b, z\rangle$ where
$i$ consists of $\lceil \log N\rceil$ bits, $b$ is one bit and
$z$ consists of all other bits. Then, $O$ maps
$|i, b, z\rangle$ maps to $|i, b\oplus x_i, z\rangle$.
(i.e., the first $\lceil\log N\rceil$ bits are interpreted as an index
$i$ for an input bit $x_i$ and this input bit is XORed on
the next qubit.) We use $O_x$ to denote the query transformation
corresponding to an input $x=(x_1, \ldots, x_n)$.

Also, we can define that $O$ maps $\ket{i, b, z}$ to 
$(-1)^{b\cdot x_i}\ket{i, b, z}$
(i.e., instead of XORing $x_i$ on an extra qubit we change phase
depending on $x_i$). It is well known that both definitions 
are equivalent up to a constant factor: one query of the $1^{\rm st}$ type 
can be simulated with a one query of the $2^{\rm nd}$ type and one query 
of the $2^{\rm nd}$ type can be simulated 
with 2 queries of the $1^{\rm st}$ type. 
For technical convenience, we use the $2^{\rm nd}$ 
definition in most of this paper.

The computation starts with a state $|0\rangle$.
Then, we apply $U_0$, $O$, $\ldots$, $O$,
$U_T$ and measure the final state.
The result of the computation is the rightmost bit of
the state obtained by the measurement.

The quantum computation computes
$f$ with bounded error if, for every $x=(x_1, \ldots, x_N)$,
the probability that the rightmost bit 
of $U_T O_x U_{T-1} \ldots O_x U_0\ket{0}$ 
equals  $f(x_1, \ldots, x_N)$ is at
least $1-\epsilon$ for some fixed $\epsilon<1/2$.

This model can be easily extended to functions defined on
a larger set (for example, $\{1, \ldots, N\}$) or functions
having more than 2 values.
In the first case, we replace one bit $b$ with several bits
($\lceil\log N\rceil$ bits in the case of $\{1, \ldots, N\}$).
In the second case, we measure several rightmost bits to obtain
the answer.

\section{The main idea}
\label{section31}

Let $S$ be a subset of the set of possible inputs $\{0, 1\}^N$.
We run the algorithm on a superposition of inputs in $S$.
More formally, let $\H_A$ be the workspace of the algorithm.
We consider a bipartite system $\H=\H_A\otimes \H_I$
where $\H_I$ is an ``input subspace" spanned by
basis vectors $\ket{x}$ corresponding to inputs $x\in S$.

Let $U_T O U_{T-1} \ldots U_0$ be the sequence of unitary transformations
on $\H_A$ performed by the algorithm $A$
(with $U_0, \ldots, U_T$ being the transformations that
do not depend on the input and $O$ being the query transformations).
We transform it into a sequence of unitary transformations on $\H$.
A unitary transformation $U_i$ on $\H_A$ corresponds to
the transformation $U'_i=U_i\otimes I$ on the whole $\H$.
The query transformation $O$ corresponds to a transformation $O'$
that is equal to $O_x$ on subspace $H_A\otimes \ket{x}$.

We perform the sequence of transformations
$U'_T O' U'_{T-1}\ldots U'_0$ on the starting state
\[ \ket{\psi_{start}}=\ket{0}\otimes \sum_{x\in S} \alpha_x \ket{x} .\]
Then, the final state is
\[ \ket{\psi_{end}}= \sum_{x\in S}\alpha_x \ket{\psi_x}\otimes\ket{x} \]
where $\ket{\psi_x}$ is the final state of $A=U_T O U_{T-1} \ldots U_0$
on the input $x$. This follows from the fact that the restrictions
of $U'_T, O', U'_{T-1}, \ldots, U'_0$ to $\H_A\otimes\ket{x}$ are
$U_T$, $O_x$, $U_{T-1}$, $\ldots$, $U_0$ and these are exactly the
transformations of the algorithm $A$ on the input $x$.

In the starting state, the $\H_A$ and $\H_I$ parts
of the superposition are unentangled. In the final state,
however, they must be entangled (if the algorithm works correctly).
To see that, consider a simple example where the algorithm
has to recover the whole input. 

Let $\alpha_x=1/\sqrt{m}$ (where $m=|S|$)
for all $x\in S$.
In the exact model (the algorithm is not allowed to give the wrong answer
even with a small probability), $\ket{\psi_x}$ must be $\ket{x}\ket{\varphi_x}$
where $\ket{x}$ is the answer of the algorithm and $\ket{\varphi_x}$ are 
algorithm's workbits. This means that the final state is
\[  \frac{1}{\sqrt{m}} \sum_{x\in S}\ket{x}\ket{\varphi_x}\otimes\ket{x}, \]
i.e., it is fully entangled state. In the bounded error model
(the algorithm can give a wrong answer with a probability at most $\epsilon$),
$\ket{\psi_x}$ must be $(1-\epsilon)\ket{x}\ket{\varphi_x}+\ket{\psi'_x}$
and the final state must be
\[  \frac{1}{\sqrt{m}} \sum_{x\in S}
((1-\epsilon)\ket{x}\ket{\varphi_x}+\ket{\psi'_x})\otimes\ket{x} \]
which is also quite highly entangled.

In the general case (we have to compute some function $f$ instead
of learning the whole input $x$), the parts of $\H_I$ corresponding
to inputs with $f(x)=z$ must become entangled with parts of $\H_A$
corresponding to the answer $z$.

Thus, we can show a lower bound on quantum query algorithms  
by showing that, given an unentangled start state, we cannot
achieve a highly entangled end state with less than a certain
number of query transformations.

Next, we describe more formally how we bound this entanglement.
If we trace out $\H_A$ from the states $\ket{\psi_{start}}$ and
$\ket{\psi_{end}}$, we obtain mixed states over $\H_I$. Let
$\rho_{start}$ and $\rho_{end}$ be the density matrices describing
these states.

$\rho_{start}$ is a $m\times m$ matrix corresponding to the
pure state $\sum_{x\in S} \alpha_x \ket{x}$.
Entries of this matrix are $(\rho_{start})_{xy}=\alpha^{*}_x \alpha_y$.
In particular, if the start state is 
$\frac{1}{\sqrt{m}} \sum_{x\in S}\ket{x}$,
all entries of the $\rho_{start}$ are $1/m$.

For $\rho_{end}$ we have

\begin{Lemma}
\label{L1}
Let $A$ be an algorithm that computes $f$ with probability at least 
$1-\epsilon$.
Let $x, y$ be such that $f(x)\neq f(y)$.
Then, 
\[ |(\rho_{end})_{xy}|\leq 2\sqrt{\epsilon(1-\epsilon)}
|\alpha_x||\alpha_y|. \] 
\end{Lemma}

\proof
Let $\ket{\psi_x}$, $\ket{\psi_y}$ be the final superpositions of the
algorithm $A$
on inputs $x, y$. We take a basis for $\H_A$ consisting of
the vectors of the form $\ket{z}\ket{v}$ where $\ket{z}$ is
a basis for the answer part (the part which is measured at the end of
algorithm to obtain the answer) and $\ket{v}$ is a basis for the rest
of $\H_A$ (workbits). We express $\ket{\psi_x}$ and $\ket{\psi_y}$ in
this basis.
Let
\[ \ket{\psi_x}=\sum_{z,v}a_{z,v}\ket{z}\ket{v}, \mbox{         }
\ket{\psi_y}=\sum_{z,v}b_{z,v}\ket{z}\ket{v} .\]
The final state of the algorithm is 
$\sum_{x\in S} \alpha_x \ket{\psi_x}\otimes\ket{x}$.
By tracing out $\H_A$ in the $\ket{z}\ket{v}$ basis,
we get 
\[ (\rho_{end})_{xy}=\alpha_x \alpha_y\sum_{z,v} a^*_{z, v} b_{z,v}. \]

Define
$\epsilon_1=\sum_{z, v: z\neq f(x)} |a_{z,v}|^2$ 
and $\epsilon_2=\sum_{z, v: z=f(x)} |b_{z,v}|^2$.
Then, $\epsilon_1\leq \epsilon$ and $\epsilon_2\leq \epsilon$ (because
these are the probabilities that the measurement at the end
of algorithm gives us a wrong answer: not $f(x)$ for the input $x$ and
$f(x)$ for the input $y$).
We have
\[ |\sum_{z,v} a^*_{z, v} b_{z,v}| \leq \sum_{z,v} |a_{z, v}| |b_{z,v}| 
 = \sum_{z,v: z=f(x)} |a_{z, v}| |b_{z,v}| + 
\sum_{z,v:z\neq f(x)} |a_{z, v}| |b_{z,v}| \]
\[ \leq  \sqrt{\sum_{z,v: z=f(x)} |a_{z, v}|^2} 
\sqrt{\sum_{z,v: z=f(x)} |b_{z, v}|^2} 
 +\sqrt{\sum_{z,v: z\neq f(x)} |a_{z, v}|^2}
\sqrt{\sum_{z,v: z\neq f(x)} |b_{z, v}|^2} 
 = \sqrt{(1-\epsilon_1)\epsilon_2}+\sqrt{\epsilon_1(1-\epsilon_2)} .\]
This expression is maximized by $\epsilon_1=\epsilon_2=\epsilon$,
giving us $2\sqrt{\epsilon(1-\epsilon)}$.
Therefore,
\[ |(\rho_{end})_{xy}|= |\alpha_x| |\alpha_y| |\sum_{z,v} a^*_{z,
v} b_{z,v}| \leq
2\sqrt{\epsilon(1-\epsilon)} |\alpha_x| |\alpha_y|. \]
\qed

In particular, if $\ket{\psi_{start}}$ is the
uniform $\frac{1}{\sqrt{m}} \sum_{x\in S} \ket{x}$,
we have $(\rho_{end})_{xy}\leq 2\sqrt{\epsilon(1-\epsilon)}/m$.
Note that, for any $\epsilon<1/2$, $2\sqrt{\epsilon(1-\epsilon)}<1$.
Thus, if the algorithm $A$ works correctly,
the absolute value of every entry of $\rho_{end}$
that corresponds to inputs $x, y$ with $f(x)\neq f(y)$
must be smaller than the corresponding entry of $\rho_{start}$
by a constant fraction.

To prove a lower bound on the number of queries, we bound
the change in $\rho_{xy}$ caused by one query.
Together with Lemma \ref{L1}, this implies a lower bound
on the number of queries.

\section{Lower bound on search}

Next, we apply this technique to several problems.
We start with the simplest case: the lower bound on
unordered search problem (Theorem \ref{Gr}).
Then, we show two general lower bound theorems (Theorems \ref{Main}
and \ref{Main1}).

Each of these theorems is a special case of the next one: 
Theorem \ref{Main} implies Theorem \ref{Gr}
and Theorem \ref{Main1} implies Theorem \ref{Main}.
However, more general theorems have more complicated proofs
and it is easier to see the main idea in the simple case
of unordered search. Therefore, we show this case first,
before general theorems \ref{Main} and \ref{Main1}.

\noindent
{\bf Problem:}
We are given $x_1, \ldots, x_N\in \{0, 1\}$ and we have to find $i$
such that $x_i=1$.

\begin{Theorem}
\label{Gr}
\cite{BBBV}
Any quantum algorithm that finds $i$ with probability $1-\epsilon$
uses $\Omega(\sqrt{N})$ queries.
\end{Theorem}

\proof
Let $S$ be the set of inputs with one $x_i$ equal to 1 and the rest 0.
Then, $|S|=N$ and $\H_I$ is an $N$-dimensional space.
To simplify the notation, we use $\ket{i}$ to denote the basis
state of $\H_I$ corresponding to the input $(x_1, \ldots, x_N)$
with $x_i=1$.

Let $\rho_k$ be the density matrix of $\H_I$
after $k$ queries.
Note that $\rho_0=\rho_{start}$ and $\rho_T=\rho_{end}$.
We consider the sum of absolute values of 
all its off-diagonal entries 
$S_k=\sum_{x, y, x\neq y} |(\rho_{k})_{xy}|$.
We will show that
\begin{enumerate}
\item
$S_0=N-1$,
\item
$S_T\leq 2\sqrt{\epsilon(1-\epsilon)}(N-1)$
\item
$S_{k-1}-S_k\leq 2\sqrt{N-1}$ for all $k\in\{1, \ldots, T\}$.
\end{enumerate}
This implies that the number of queries $T$ is at least
$(1-2\sqrt{\epsilon(1-\epsilon)})\sqrt{N-1}/2$.

The first two properties follow straightforwardly from the results
at the end of section \ref{section31}. 
$N\times N$ matrices $\rho_i$ have $N(N-1)$ non-diagonal entries and 
each of these entries is $1/N$ in $\rho_{start}$
and at most $2\sqrt{\epsilon(1-\epsilon)}/N$ 
in $\rho_{end}$ (Lemma \ref{L1} together
with the fact that each of these entries corresponds to two
inputs with different answers). 

It remains to prove the third part.
First, notice that
\[ S_{k-1}-S_k =\sum_{x,y:x\neq y} |(\rho_{k-1})_{xy}| -
\sum_{x,y:x\neq y} |(\rho_k)_{xy}| \leq
\sum_{x, y:x\neq y}|(\rho_{k-1})_{xy}-(\rho_k)_{xy}| .\]
Therefore, it suffices to bound the sum of 
$|(\rho_{k-1})_{xy}-(\rho_k)_{xy}|$.
%

A query corresponds to representing the pure state before the
query as
\[ \ket{\psi_{k-1}} = \sum_{i, z} \sqrt{p_{i, z}} 
\ket{i, z} \otimes \ket{\psi_{i, z}} ,\]
\[ \ket{\psi_{i, z}}=\sum_{j=1}^n \alpha_{i, z, j}\ket{j} \]
and changing the phase on the $\ket{i}$ component of $\ket{\psi_{i,
z}}$. If we consider just the $H_I$ part, the density matrix
$\rho_{k-1}$ before the query is $\sum_{i, z}p_{i, z}\ket{\psi_{i,
z}}\bra{\psi_{i, z}}$. The density matrix $\rho_k$ after the query
is $\sum_{i, z}p_{i, z}\ket{\psi'_{i, z}}\bra{\psi'_{i, z}}$ where
\[ \ket{\psi'_{i, z}}= \sum_{j\neq i}\alpha_{i, z, j}\ket{j}-\alpha_{i,
z, i}\ket{i}.\]

Consider $\rho_{i, z}=\ket{\psi_{i, z}}\bra{\psi_{i, z}}$ and
$\rho'_{i, z}=\ket{\psi'_{i, z}}\bra{\psi'_{i, z}}$. 
Then, $\rho_{k-1}=\sum_{i,z} p_{i,z}\rho_{i,z}$ and
$\rho_{k}=\sum_{i,z} p_{i,z}\rho'_{i,z}$.

The only
entries where $\rho_{i,z}$ and $\rho'_{i,z}$ differ are 
the entries in the $i^{\rm th}$
column and the $i^{\rm th}$ row. These entries are $\alpha^*_{i, z,
j}\alpha_{i, z, i}$ in $\rho_{i, z}$ and -$\alpha^*_{i, z,
j}\alpha_{i, z, i}$ in $\rho'_{i, z}$. The sum of absolute values
of the differences of all entries in the $i^{\rm th}$ row is
\[ \sum_{j\neq i} 2|\alpha^*_{i, z, j}\alpha_{i, z, i}|\leq
2 |\alpha_{i, z, i}| \sum_{j\neq i} |\alpha_{i, z, j}| .\] 
Similarly, the sum of absolute values
of the differences of all entries in the $i^{\rm th}$ column is
at most $2 |\alpha_{i, z, i}| \sum_{j\neq i} |\alpha_{i, z, j}|$ as well.
So, the sum of absolute values of all differences is 
at most $4 |\alpha_{i, z, i}| \sum_{j\neq i} |\alpha_{i, z, j}|$.

By Cauchy-Schwartz inequality,
\[ \sum_{j\neq i} |\alpha_{i, z, j}|\leq \sqrt{N-1}
\sqrt{\sum_{j\neq i}|\alpha_{i, z, j}|^2} 
= \sqrt{N-1}
\sqrt{1-|\alpha_{i, z, i}|^2} .\] Therefore,
\[ \sum_{j\neq i} 4|\alpha^*_{i, z, j}\alpha_{i, z, i}|\leq
4 \sqrt{N-1} |\alpha_{i, z, i}|\sqrt{1-|\alpha_{i, z, i}|^2}\leq 2
\sqrt{N-1}.\]

Define $S_{i,z}=\sum_{x,y:x\neq y} |(\rho_{i,z})_{xy}-(\rho'_{i,z})_{xy}|$.
Then, we have just shown $S_{i,z}\leq 2\sqrt{N-1}$.
This implies a bound on the sum 
$\sum_{x,y:x\neq y}|(\rho_{k-1})_{xy}-(\rho_k)_{xy}|$.

\[ \sum_{x,y:x\neq y}|(\rho_{k-1})_{xy}-(\rho_k)_{xy}| 
= \sum_{x,y:x\neq y} |\sum_{i,z} p_{i,z} (\rho_{i,z})_{xy}-
\sum_{i,z} p_{i,z}(\rho'_{i,z})_{xy}| \]
\[ \leq \sum_{i,z} p_{i,z} \sum_{x,y:x\neq y}
|(\rho_{i,z})_{xy}-(\rho'_{i,z})_{xy}| 
\leq \sum_{i,z} p_{i,z} 2\sqrt{N-1}=2\sqrt{N-1} .\]
This completes the proof. \qed

\section{The general lower bound}

\subsection{The result}

Next, we obtain a general lower bound theorem.

\begin{Theorem}
\label{Main}
Let $f(x_1, \ldots, x_N)$ be a function of $n$ $\{0, 1\}$-valued
variables and $X, Y$ be two sets of inputs such that $f(x)\neq f(y)$
if $x\in X$ and $y\in Y$.
Let $R\subset X \times Y$ be such that
\begin{enumerate}
\item
For every $x\in X$, there exist at least $m$ different $y\in Y$ such that
$(x, y)\in R$.
\item
For every $y\in Y$, there exist at least $m'$ different $x\in X$ such that
$(x, y)\in R$.
\item
For every $x\in X$ and $i\in\{1, \ldots, n\}$, there are at most $l$ 
different $y\in Y$ such that $(x, y)\in R$ and $x_i\neq y_i$.
\item
For every $y\in Y$ and $i\in\{1, \ldots, n\}$, there are at most $l'$ 
different $x\in X$ such that $(x, y)\in R$ and $x_i\neq y_i$.
\end{enumerate}
Then, any quantum algorithm computing $f$ uses  
$\Omega(\sqrt{\frac{m m'}{l l'}})$ queries.
\end{Theorem} 

\proof
Consider the set of inputs $S=X\cup Y$ and the superposition 
\[ \frac{1}{\sqrt{2|X|}}\sum_{x\in X} \ket{x}+ 
\frac{1}{\sqrt{2|Y|}}\sum_{y\in Y} \ket{y}\] 
over this set of inputs.
Let $S_i$ be the sum of $|(\rho_i)_{xy}|$ over all $x, y$ such that
$(x, y)\in R$. Then, the theorem follows from
\begin{enumerate}
\item
$S_0-S_T\geq (1-2\sqrt{\epsilon(1-\epsilon)}) \sqrt{mm'}$
\item
$S_{k-1}-S_k\leq \sqrt{l l'}$
\end{enumerate}

To show the first part, let $(x, y)\in R$. 
Then, $(\rho_0)_{xy}=\frac{1}{\sqrt{|X||Y|}}$ and $|(\rho_T)_{xy}|\leq 
\frac{2\sqrt{\epsilon(1-\epsilon)}}{\sqrt{|X||Y|}}$ (Lemma \ref{L1}).
Therefore, 
\[ |(\rho_0)_{xy}|-|(\rho_T)_{xy}| \geq 
\frac{1-2\sqrt{\epsilon(1-\epsilon)}}{\sqrt{|X||Y|}} .\]
The number of $(x, y)\in R$ is at least $\max(|X|m, |Y|m')$
because for every $x\in X$, there are at least $m$ possible $y\in Y$ and,
for every $y\in Y$, there are at least $m'$ possible $x\in X$.
We have
\[ \max(|X|m, |Y|m')\geq \frac{|X|m+|Y|m'}{2} \geq \sqrt{|X||Y|m m'} ,\]
\[ S_0-S_T \geq \sqrt{|X||Y|m m'}
\frac{1-2\sqrt{\epsilon(1-\epsilon)}}{\sqrt{|X||Y|}} 
=(1-2\sqrt{\epsilon(1-\epsilon)})\sqrt{m m'} .\]

Next, we prove the second part.
Similarly to the previous proof, we represent
\[ \ket{\psi_{k-1}} = \sum_{i, z} \sqrt{p_{i, z}} 
\ket{i, z} \otimes \ket{\psi_{i, z}} ,\]
\[ \ket{\psi_{i, z}}=\sum_{x\in S} \alpha_{i, z, x}\ket{x} .\]
A query corresponds to changing the sign on all components
with $x_i=1$. It transforms $\ket{\psi_{i, z}}$ to
\[ \ket{\psi'_{i, z}}=\sum_{x\in S: x_i=0} \alpha_{i, z, x}\ket{x}-
 \sum_{x\in S: x_i=1} \alpha_{i, z, x}\ket{x} .\]
Let $\rho_{i, z}=\ket{\psi_{i, z}}\bra{\psi_{i, z}}$ and
$\rho'_{i, z}=\ket{\psi'_{i, z}}\bra{\psi'_{i, z}}$.
We define $S_{i, z}=\sum_{(x, y)\in R} |(\rho_{i, z})_{xy}-
(\rho'_{i, z})_{xy}|$. 

If $x_i=y_i$, then $(\rho_{i, z})_{xy}$ and $(\rho'_{i, z})_{xy}$
are the same. If one of $x_i$, $y_i$ is 0 and the other is 1,
$(\rho_{i, z})_{xy}=-(\rho'_{i, z})_{xy}$ and
$|(\rho_{i, z})_{xy}- (\rho'_{i, z})_{xy}|=2 |(\rho_{i, z})_{xy}|
=2|\alpha_{i,z,x}||\alpha_{i,z,y}|$.
Therefore,
\[ S_{i,z}=\sum_{(x, y)\in R:x_i\neq y_i} 2|\alpha_{i,z,x}||\alpha_{i,z,y}| 
\leq \sum_{(x, y)\in R:x_i\neq y_i} \sqrt{\frac{l'}{l}} |\alpha_{i,z,x}|^2
+\sqrt{\frac{l}{l'}} |\alpha_{i,z,y}|^2 \]
\[ \leq \sum_{x\in X} l \sqrt{\frac{l'}{l}} |\alpha_{i,z,x}|^2 +
 \sum_{y\in Y} l' \sqrt{\frac{l}{l'}} |\alpha_{i,z,y}|^2 
= \sqrt{l l'} \sum_{x\in X\cup Y} |\alpha_{i,z,x}|^2 =\sqrt{l l'} .\]
Similarly to the previous proof, this implies the same bound on 
$S_{k-1}-S_k$.
\qed

\subsection{Relation to block sensitivity bound}

Our Theorem \ref{Main} generalizes the block 
sensitivity bound of \cite{Beals,V}.

Let $f$ be a Boolean function and $x=(x_1, \ldots, x_n)$ an input to $f$.
For a set $S\subseteq\{1, \ldots, n\}$, $x^{(S)}$ denotes
the input obtained from $x$ by flipping all variables $x_i$, $i\in S$.
$f$ is {\em sensitive} to $S$ on input $x$ if
$f(x)\neq f(x^{(S)})$.

The block sensitivity of $f$ on input $x$ is the maximal number $t$
such that there exist $t$ pairwise disjoint sets $S_1$, $S_2$, $\ldots$, $S_t$
such that, for all $i\in\{1, \ldots, t\}$, $f$ is sensitive to $S_i$ on $x$.
We denote it by $bs_x(f)$.
The block sensitivity of $f$, $bs(f)$ is just the maximum of $bs_x(f)$
over all inputs $x$\cite{Nisan}.

\begin{Theorem}
\cite{Beals,V}
Let $f$ be any Boolean function. Then, any quantum query algorithm computing
$f$ uses $\Omega(\sqrt{bs(f)})$ queries.
\end{Theorem}

To see that this is a particular case of Theorem \ref{Main},
let $x$ be the input on which $f$ achieves $bs(f)$ block sensitivity.
Then, we can take $X=\{x\}$ and $Y=\{x^{(S_1)}, \ldots, x^{(S_{bs(f)})}\}$.

Let $R=\{(x, x^{(S_1)}), (x, x^{(S_2)}), \ldots, (x, x^{(S_{bs(f)})})\}$.
Then, $m=bs(f)$, $m'=1$. Also, $l=1$ (because, by the definition of the
block sensitivity, $m$ blocks of input variables have to be
disjoint) and $l'=1$.
Therefore, we get $\frac{m m'}{l l'}=bs(f)$ and 
Theorem \ref{Main} gives the $\Omega(\sqrt{bs(f)})$ lower bound
for any Boolean function $f$. 

In the next subsection we show some problems for which our method 
gives a better bound than the block-sensitivity method.

\subsection{Applications}

For a first application, consider AND of ORs:
\[ f(x_1, \ldots, x_N)= (x_1 OR x_2
\ldots OR x_{\sqrt{N}}) AND \] \[ (x_{\sqrt{N}+1} \ldots x_{2\sqrt{N}})
 AND \ldots AND (x_{N-\sqrt{N}+1} OR \ldots OR x_{N})\] where $x_1,
\ldots, x_N\in\{0, 1\}$.

$f$ can be computed with $O(\sqrt{N}\log N)$ queries by a
two-level version of Grover's algorithm (see \cite{BCW}). However,
a straightforward application of lower bound methods from
\cite{BBBV,Beals} only gives an $\Omega(\sqrt[4]{N})$ bound because the
block sensitivity of $f$ is $\Theta(\sqrt{N})$ and the lower bound
on the number of queries given by hybrid or polynomials
method is the square root of block sensitivity\cite{Beals,V}.
Our method gives

\begin{Theorem}
Any quantum algorithm computing AND of ORs uses $\Omega(\sqrt{N})$
queries.
\end{Theorem}

\proof
For this problem, let $X$ be the set of all
$x=(x_1, \ldots, x_n)$ such that, for every $i\in\{1, \ldots, \sqrt{N}\}$, 
there is exactly one $j\in\{1, \ldots, \sqrt{n}\}$ with $x_{(i-1)\sqrt{N}+j}=1$.
$Y$ is the set of all $y=(y_1, \ldots, y_n)$ such that
$y_{(i-1)\sqrt{N}+1}=\ldots = y_{i\sqrt{N}}=0$ for some $i$ and, 
for every $i'\neq i$, there is a unique
$j\in\{1, \ldots, \sqrt{N}\}$ with $y_{(i-1)\sqrt{N}+j}=1$.
$R$ consists of all pairs $(x, y)$, $x\in X$, $y\in Y$ such that
there is exactly one $i$ with $x_i\neq y_i$.

Then, $m=m'=\sqrt{n}$ because, given $x\in X$, there are $\sqrt{n}$
1s that can be replaced by 0 and replacing any one of them gives
some $y\in Y$. Conversely, if $y\in Y$, there are $\sqrt{n}$ ways to
add one more 1 so that we get $x\in X$. 
On the other hand, $l=l'=1$ because, given $x\in X$ (or $y\in Y$)
and $i\in\{1, \ldots, n\}$, there is only one input that differs
from $x$ only in the $i^{\rm th}$ position.
Therefore, $\sqrt{\frac{m m'}{l l'}}=\sqrt{n}$
and the result follows from Theorem \ref{Main}.
\qed

Our theorem can be also used to give another proof for the following theorem
of Nayak and Wu\cite{NW}.

\begin{Theorem}
\label{NWT}
\cite{NW}
Let $f:\{0, 1, \ldots, n-1\}\rightarrow \{0, 1\}$ be a Boolean function
that is equal to 1 either at exactly $n/2$ points of the domain
or at exactly $(1+\epsilon)n/2$ points.
Then, any quantum algorithm that determines whether 
the number of points where $f(x)=1$ is $n/2$ or $(1+\epsilon)n/2$
uses $\Omega(\frac{1}{\epsilon})$ queries.
\end{Theorem}

This result implies lower bounds on the number of quantum
queries needed to compute (or to approximate) 
the median of $n$ numbers\cite{NW}. 
It was shown in \cite{NW} using polynomials method.
No proof that uses adversary arguments similar to \cite{BBBV,V} is known.

With our ``quantum adversary" method, Theorem \ref{NWT} can be proven
in similarly to other theorems in this paper.

\proof
Let $X$ be the set of all $f$ that are 1 at exactly $n/2$ points,
$Y$ be the set of all $f$ that are 1 at $(1+\epsilon)n/2$ points
and $R$ be the set of all $(f, f')$ such that $f\in X$, $f'\in Y$
and they differ in exactly $\epsilon n/2$ points.

Then, $m={ n/2 \choose \epsilon n/2}$ and $m'={(1+\epsilon) n/2 \choose 
\epsilon n/2}$. On the other hand, 
$l={n/2-1 \choose \epsilon n/2 -1}$ and $l'={(1+\epsilon) n/2 -1 \choose
\epsilon n/2-1}$.
Therefore,
\[ \frac{m m'}{l l'} = \frac{ {n/2 \choose \epsilon n/2}
{(1+\epsilon) n/2 \choose \epsilon n/2} }{ {n/2-1 \choose \epsilon n/2 -1}
{(1+\epsilon) n/2 -1 \choose \epsilon n/2-1} } 
=\frac{\frac{n}{2} \frac{(1+\epsilon)n}{2}}
{\frac{\epsilon n}{2}\frac{\epsilon n}{2}} =
\frac{1+\epsilon}{\epsilon^2}>\frac{1}{\epsilon^2} .\]
By Theorem \ref{Main}, the number of queries is
$\Omega(\sqrt{\frac{m m'}{l l'}})=\Omega(1/\epsilon)$.

\qed

There are several other known lower bounds that also follow from
Theorem \ref{Main}. In particular, Theorem \ref{Main} implies 
$\Omega(N)$ lower bounds for MAJORITY and PARITY of \cite{Beals}.

\section{Inverting a permutation}

\subsection{Extension of Theorem 2}

For some lower bounds (like inverting a permutation), 
we need a following extension of Theorem \ref{Main}.
This is our most general result.

\begin{Theorem}
\label{Main1}
Let $f(x_1, \ldots, x_N)$ be a function of $n$ variables 
with values from some finite set
and $X, Y$ be two sets of inputs such that $f(x)\neq f(y)$
if $x\in X$ and $y\in Y$.
Let $R\subset X \times Y$ be such that
\begin{enumerate}
\item
For every $x\in X$, there exist at least $m$ different $y\in Y$ such that
$(x, y)\in R$.
\item
For every $y\in Y$, there exist at least $m'$ different $x\in X$ such that
$(x, y)\in R$.
\end{enumerate}
Let $l_{x, i}$ be the number of $y\in Y$ such that $(x, y)\in R$ and $x_i\neq y_i$
and $l_{y, i}$ be the number of $x\in X$ such that $(x, y)\in R$ and $x_i\neq y_i$.
Let $l_{max}$ be the maximum of $l_{x, i}l_{y, i}$ over all $(x, y)\in R$
and $i\in\{1, \ldots, N\}$ such that $x_i\neq y_i$.
Then, any quantum algorithm computing $f$ uses  
$\Omega(\sqrt{\frac{m m'}{l_{max}}})$ queries.
\end{Theorem}

The parameters $l$ and $l'$ of Theorem \ref{Main} are just
$\max_{x\in X, i} l_{x, i}$ and $\max_{y\in Y, i} l_{y, i}$.
It is easy to see that
\[ \max_{(x, y)\in R, x_i\neq y_i} l_{x, i} l_{y, i} 
\leq \max_{x, i} l_{x, i} \max_{y, i} l_{y, i} .\]
Therefore, the lower bound
given by Theorem \ref{Main1} is always greater than or equal to the lower
bound of Theorem \ref{Main}. 
However, Theorem \ref{Main1} gives a better bound
if, for every $(x, y)\in R$ and $i$, at least one of $l_{x, i}$ or $l_{y, i}$ 
is less than its maximal value (which happens for inverting a permutation).

Also, Theorem \ref{Main1} allows $\{1, \ldots, N\}$-valued
variables instead of only $\{0, 1\}$-valued in Theorem \ref{Main}.

\proof
Similarly to Theorem \ref{Main1}, we consider 
the set of inputs $S=X\cup Y$ and the superposition 
\[ \frac{1}{\sqrt{2|X|}}\sum_{x\in X} \ket{x}+ 
\frac{1}{\sqrt{2|Y|}}\sum_{y\in Y} \ket{y}\] 
over this set of inputs.
Let $S_k$ be the sum of $|(\rho_k)_{xy}|$ over all $x, y$ such that
$(x, y)\in R$. The theorem follows from
\begin{enumerate}
\item
$S_T-S_0\geq (1-2\sqrt{\epsilon(1-\epsilon)}) \sqrt{mm'}$
\item
$S_{k-1}-S_k\leq \sqrt{l_{max}}$
\end{enumerate}

The first part is shown in the same way as in the proof of Theorem \ref{Main}.
For the second part, express the state before the $k^{\rm th}$
query as
\[ \ket{\psi_{k-1}} = \sum_{i, a, z,x} \alpha_{i,a,z,x} \ket{i, a, z}
\otimes \ket{x} \]
where $i$ is the index of the input variable $x_i$ being
queried, $a$ are $\log N$ bits for the answer, $z$ is the part of $\H_A$
that does not participate in the query (extra workbits) and 
$x$ is $\H_I$ part of the superposition.
A query changes this to
\[ \ket{\psi_k}= \sum_{i, a, z,x} \alpha_{i,a,z,x} \ket{i, a\oplus x_i, z}
\otimes \ket{x} 
=\sum_{i, a, z,x}\alpha_{i,a\oplus x_i,z,x} \ket{i, a, z}
\otimes \ket{x} .\]
Denote
\[ \ket{\psi_{i,a,z}}=\sum_x \alpha_{i,a,z,x} \ket{x}, \mbox{   }
\ket{\psi'_{i,a,z}}=\sum_x \alpha_{i,a\oplus x_i,z,x} \ket{x}.\]
$\rho_{k-1, i}=\sum_{a,z} \ket{\psi_{i, a, z}}\bra{\psi_{i, a, z}}$ 
and $\rho_{k, i} =\sum_{a,z} \ket{\psi'_{i, a, z}}\bra{\psi'_{i, a, z}}$ 
are the parts of $\rho_{k-1}$ and $\rho_k$ corresponding to querying $i$.
We have $\rho_{k-1}=\sum_{i=1}^n \rho_{k-1, i}$ and
$\rho_{k}=\sum_{i=1}^n \rho_{k, i}$.

Let $S_{k, i}$ be sum of absolute values of differences
$|(\rho_{k-1,i})_{xy}-(\rho_{k,i})_{xy}|$ over all
$(x, y)\in R$.
Then, for every $x, y$,
\[ |(\rho_{k-1})_{xy}|-|(\rho_{k})_{xy}| \leq
|(\rho_{k-1})_{xy}-(\rho_k)_{xy}| 
= | \sum_{i} (\rho_{k-1,i})_{xy} - \sum_{i} (\rho_{k,i})_{xy} | \leq
\sum_i |(\rho_{k-1,i})_{xy}-(\rho_{k,i})_{xy}| .\]
Therefore (by summing over all such $x$ and $y$),
$S_{k-1}-S_k \leq \sum_i S_{k,i}$ and we can bound  
$S_{k-1}-S_k$ by bounding $S_{k,i}$.

Let $x$, $y$ be two inputs such that $x_i=y_i$. Then, it is
easy to see that
\[ (\rho_{k-1, i})_{xy}=\sum_{a, z}\alpha^*_{i, a, z, x}\alpha_{i, a, z, y}
=\sum_{a, z} \alpha^*_{i, a\oplus x_i, z, x}\alpha_{i, a\oplus y_i, z, y}=
(\rho_{k, i})_{xy}.\] Therefore, the only non-zero entries 
in $S_{k, i}$ are the entries corresponding to
$(x, y)\in R$ with $x_i\neq y_i$. The sum of 
their differences $|(\rho_{k-1,i})_{xy}- (\rho_{k,i})_{xy}|$
is at most the sum of absolute values of such
entries in $\rho_{k-1, i}$ plus the sum of absolute values of them in
$\rho_{k, i}$. We bound these two sums.

First, any density matrix is semipositive definite.
This implies that 
\[ |(\rho_{k-1, i})_{xy}| \leq 
\frac{1}{2}\left(\sqrt{\frac{l_{y,i}}{l_{x,i}}} |(\rho_{k-1, i})_{xx}|+
\sqrt{\frac{l_{x,i}}{l_{y,i}}} |(\rho_{k-1, i})_{yy}|\right).\]
for any $x$ and $y$. Therefore,
\[ \mathop{\sum_{x, y:(x, y)\in R}}_{x_i\neq y_i}
|(\rho_{k-1, i})_{xy}| 
\leq \frac{1}{2} \mathop{\sum_{x, y:(x, y)\in R}}_{x_i\neq y_i}
\sqrt{\frac{l_{y,i}}{l_{x,i}}} |(\rho_{k-1, i})_{xx}|+
\sqrt{\frac{l_{x,i}}{l_{y,i}}} |(\rho_{k-1, i})_{yy}|
=\frac{1}{2} \sum_{x\in X\cup Y} l_{x,i} \sqrt{\frac{l_{y,i}}{l_{x,i}}}  
|(\rho_{k-1, i})_{xx}|\]
\[ = \frac{1}{2} \sum_{x\in X\cup Y} 
\sqrt{l_{x, i} l_{y, i}} |(\rho_{k-1, i})_{xx}| 
\leq \frac{1}{2}\sqrt{l_{max}}\sum_{x\in X\cup Y}  
|(\rho_{k-1, i})_{xx}| =
\frac{\sqrt{l_{max}}}{2} Tr \rho_{k-1, i} .\]
The same argument shows that a similar sum
is at most $\frac{\sqrt{l_{max}}}{2} Tr\rho_{k, i}$ for the matrix
$\rho_{k, i}$. Therefore,
\[ S_{k-1}-S_k\leq \sum_{i} S_{k,i} \leq
\sum_{i} \frac{\sqrt{l_{max}}}{2} (Tr\rho_{k-1, i}+Tr\rho_{k, i}) 
= \frac{\sqrt{l_{max}}}{2} (Tr\rho_{k-1}+Tr\rho_{k}) 
= \sqrt{l_{max}} .\] This
completes the proof.
\qed

\subsection{Application}

We use Theorem \ref{Main1} to prove a lower bound for
inverting a permutation\cite{BBBV}.

\noindent {\bf Problem:} We are given $x_1, \ldots, x_N\in\{1,
\ldots, N\}$ such that $(x_1, \ldots, x_N)$ is a permutation of
$\{1, \ldots, N\}$. We have to find the $i$ such that
$x_i=1$.

This problem was used in \cite{BBBV} to show
$NP^A \cap co-NP^A\not\subseteq BQP^A$ for an oracle $A$.
It is easy to see that it can be solved by Grover's algorithm
(search for $i$ with $x_i=1$). This takes $O(\sqrt{N})$
queries.

However, the $\Omega(\sqrt{N})$ lower bound proof for search problem
from \cite{BBBV} does not work for this problem.
\cite{BBBV} showed a weaker $\Omega(\sqrt[3]{N})$ bound with
a more complicated proof.

\begin{Theorem}
\label{Th2} Any quantum query algorithm that inverts a permutation
with probability $1-\epsilon$ uses $\Omega(\sqrt{N})$ queries.
\end{Theorem}

\proof 
Let $X$ be the set of all permutations $x$ with $x_i=1$ for an even $i$
and $Y$ be the set of all permutations $y$ with $y_i=1$ for an odd $i$.
$(x, y)\in R$ if $x=(x_1, \ldots, x_N)$, $y=(y_1, \ldots,
y_N)$ and there are $i, j$, $i\neq j$ such that
$x_i=y_j=1$, $x_j=y_i$ and all other elements of $x$ and $y$ are
the same.

For every $x\in X$, there are $m=n/2$ $y$ with $(x, y)\in R$.
Similarly, for every $y\in Y$, there are 
$m'=n/2$ $x$ such that $(x, y)\in R$.

Finally, if we take a pair $(x, y)\in R$ and a location $i$ such that
$x_i\neq y_i$, then one of $x_i$, $y_i$ is 1. We assume that $x_i=1$.
(The other case is similar.) Then, there are $n/2$ $y'$ such that
$(x, y')\in R$ and $x_i\neq y'_i$. However, the only $x'$ such that
$(x', y)\in R$ and $x'_i\neq y_i$ is $x'=x$. 
(Any $x'$ such that $(x', y)\in R$ and $x'_i\neq y'_i$ must also
have $x'_j\neq y_j$ where $j$ is the variable for which $y_j=1$
and $x$ is the only permutation that differs from $y$
only in these two places.)

Therefore, $l_{x, i}=n/2$, $l_{y, i}=1$ and $l_{max}=n/2$.
By Theorem \ref{Main1}, this implies that any quantum algorithm needs
$\Omega(\sqrt{\frac{n^2}{n}})=\Omega(\sqrt{n})$ queries.
\qed

\section{Relation to Grover's proof}
\label{S7}

Grover\cite{Grover1} presents a proof of the $\Omega(\sqrt{n})$ lower bound
on the search problem based on considering the sum of distances
\[ \Delta(t)=\sum_{i, j\in\{1, \ldots, n\}, i\neq j} \|\phi_i^t-\phi_0^t\|^2 \]
where $\phi_i^t$ is the state of the algorithm after $t$ queries
on the input $x_1=\ldots=x_{i-1}=0$, $x_i=1$, $x_{i+1}=\ldots=x_n=1$
and $\phi_0^t$ is the state of the algorithm after $t$ queries on
the input $x_1=\ldots=x_n=0$. 

Grover shows that, after $t$ queries, $\Delta(t)\leq 4t^2$.
If the algorithm outputs the correct answer with probability 1,
the final vectors $\phi_1^t$, $\ldots$, $\phi_n^t$ have to be
orthogonal, implying that $\Delta(t)\geq 2N-2\sqrt{N}$
(cf. \cite{Grover1}). This implies that the number of queries
must be $\Omega(\sqrt{N})$.

A similar idea (bounding a certain sum of distances) has been
also used by Shi\cite{Shi} to prove lower bounds on the number of
quantum queries in terms of average sensitivity.

These ``distance-based'' ideas can be generalized to obtain another proof of
our Theorems \ref{Main} and \ref{Main1}. Namely, for Theorem \ref{Main},
one can take
\[ \Delta(t)=\sum_{(x, y)\in R} \|\phi_x^t-\phi_y^t\|^2 \]
where $\phi_x^t$, $\phi_y^t$ are the states of the algorithm after $t$ steps on
the inputs $x$ and $y$.
Then,
\[ \|\phi_x^t-\phi_y^t\|^2 = 1 -\lbra \phi_x^t | \phi_y^t\rket^2 .\]
Let $\rho_t$ be the density matrix of $\H_I$ after $t$ steps.
By writing out the expressions for $\lbra \phi_x^t | \phi_y^t\rket$
and $(\rho_t)_{xy}$, we can see that
\[ (\rho_t)_{xy} = \frac{1}{4|X||Y|} \lbra \phi_x^t | \phi_y^t\rket .\]
This shows that the two quantities (the sum of entries in the density matrix
and the sum of distances) are quite similar. 
Indeed, we can give proofs for Theorems \ref{Main} and \ref{Main1} in terms
of distances and their sums $\Delta(t)$.
(Namely, $\Delta(0)=0$ before the first query,
$\Delta(T)$ should be large if the algorithm solves the problem with
$T$ queries and we can bound the difference $\Delta(t)-\Delta(t-1)$.
This gives the same bounds as bounding the entries of density matrices.) 

Thus, Theorems \ref{Main} and \ref{Main1} have two proofs that
are quite similar algebraically but come from two completely different
sources: running a quantum algorithm with a superposition of
inputs (our ``quantum adversary'') and looking at it from 
a geometric viewpoint (sum of distances).

The ``quantum adversary'' approach may be more general because
one could bound other quantities (besides the sum of entries in
the density matrix) which have no simple geometric interpretation.

\section{Conclusion and open problems}

We introduced a new method for proving lower bounds on quantum algorithms
and used it to prove tight (up to a multiplicative or logarithmic factor)
lower bounds on Grover's search and 3 other related problems.
Two of these bounds (Grover's search and distinguishing between an input with 
1/2 of values equal to 1 and $1/2+\epsilon$ values equal to 1) were
known before. For two other problems (inverting a permutation and 
AND of ORs), only weaker bounds were known.
One advantage of our method is that it allows to
prove all 4 bounds in a similar way. (Previous methods were
quite different for different problems.)

Some open problems:
\begin{enumerate}
\item
{\bf Collision problem\cite{Collision}.}

We are given a function 
$f:\{1, \ldots, n\}\rightarrow \{1, \ldots, n/2\}$ 
and have to find $i$, $j$ such that $f(i)=f(j)$.
Classically, this can be done by querying $f(x)$ 
for $O(\sqrt{n})$ random values
of $x$ and it is easy to see that this is optimal.
There is a quantum algorithm that solves this problem with 
$O(\sqrt[3]{n})$ queries\cite{Collision}. 
However, there is no quantum lower bound at all for this problem
(except the trivial bound of $\Omega(1)$).

The collision problem is an abstraction for {\em collision-resistant
hash functions}. If it can be solved with $O(\log n)$ queries,
there no hash function is collision-resistant against
quantum algorithms.

The exact argument that we gave in this paper (with bounding a subset of
the entries in the density matrix) does not carry over to the collision
problem. However, it may be possible to use our idea of running the algorithm
with a superposition of oracles together with some other way of measuring
the entanglement between the algorithm and the oracle.

\item
{\bf Simpler/better lower bound for binary search.}

It may be possible to simplify other lower bounds proven 
previously by different methods. 
In some cases, it is quite easy to
reprove the result by our method (like Theorem \ref{NWT})
but there are two cases in which we could not do that.
The first is the bound of \cite{Error} on the number of queries needed 
to achieve very small probability of error in database search problem. 
The second is the lower bound on the ordered search\cite{Ambainis}.
It seems unlikely that our technique can be useful in the first 
case but there is a chance that some variant of our idea 
may work for ordered search (achieving both simpler proof and
better constant under big-$\Omega$).

\item
{\bf Communication complexity of disjointness.}

Quantum communication complexity is often related to
query complexity\cite{BCW}. 
Can one use our method (either ``quantum adversary'' or distance-based
formulation) to prove lower bounds on quantum communication complexity?

A particularly interesting open problem in quantum communication
complexity is {\em set disjointness}.
The classical (both deterministic and probabilistic) 
communication complexity of set disjointness is $\Omega(n)$\cite{KS,Razborov}.
There is a quantum protocol (based on Grover's search algorithm)
that computes set disjointness with an 
$O(\sqrt{n}\log n)$ communication\cite{BCW}
but the best lower bound is only $\Omega(\log n)$\cite{BW,Ambainis1}.
\end{enumerate}

{\bf Acknowledgements.}
I would like to thank Dorit Aha\-ro\-nov, Daniel Gottesman, 
Ashwin Nayak, Umesh Vazirani and Ronald de Wolf for useful comments.

\end{document}